\documentclass[12pt,superscriptaddress,aps,prd,preprint]{revtex4}
\usepackage[active]{srcltx}
\usepackage{graphicx}
\usepackage{amsmath}
\usepackage[utf8]{inputenc}

\begin{document}
\title{1/N Expansion for Horava-Lifshitz like four-fermion models}

\author{M. Gomes}
\affiliation{Instituto de F\'\i sica, Universidade de S\~ao Paulo\\
Caixa Postal 66318, 05315-970, S\~ao Paulo, SP, Brazil}
\email{mgomes,ajsilva@if.usp.br}

\author{T. Mariz}
\affiliation{Instituto de F\'\i sica, Universidade Federal de Alagoas\\ 
57072-270, Macei\'o, Alagoas, Brazil}
\email{tmariz@fis.ufal.br}

\author{J. R. Nascimento}
\affiliation{Departamento de F\'{\i}sica, Universidade Federal da Para\'{\i}ba\\
Caixa Postal 5008, 58051-970, Jo\~ao Pessoa, Para\'{\i}ba, Brazil}
\email{jroberto,petrov@fisica.ufpb.br}

\author{A. Yu. Petrov}
\affiliation{Departamento de F\'{\i}sica, Universidade Federal da Para\'{\i}ba\\
Caixa Postal 5008, 58051-970, Jo\~ao Pessoa, Para\'{\i}ba, Brazil}
\email{jroberto,petrov@fisica.ufpb.br}

\author{A. J. da Silva}
\affiliation{Instituto de F\'\i sica, Universidade de S\~ao Paulo\\
Caixa Postal 66318, 05315-970, S\~ao Paulo, SP, Brazil}
\email{mgomes,ajsilva@if.usp.br}

\date{\today}
\begin{abstract}
We study a class of four-fermion Gross-Neveu like models in four dimensions with critical exponents $z=2$ and $z=3$. The models with $z=2$ are known to be perturbatively nonrenormalizable  but are shown to be renormalizable in the context of the $1/N$ expansion.  We calculate explicitly the effective potential for these models.
\end{abstract}
\maketitle

Horava-Lifshitz like field theories are distinguished for the presence of  higher spatial derivative terms  in the Lagrangian density, while the  terms involving temporal derivatives have the same form as in the usual relativistic theories \cite{Lifshitz,Horava,Anselmi}. In this way the canonical structure is preserved avoiding a possible unitarity breakdown. Asymptotically,  when  terms involving dimensional parameters may be neglected, these theories possess an  anisotropic scale invariance with $t\rightarrow \lambda^{-z} t $ and  $\vec x \rightarrow \lambda^{-1}\vec x$ where $z$, the so called dynamical critical exponent, measures the highest degree of the spatial derivatives. Because of the  higher derivatives, the ultraviolet behavior of the Green functions is improved allowing the completion of theories that otherwise would be nonrenormalizable. Indeed, it has been argued that gravitation would be power counting renormalizable for $z=3$ \cite{Horava}. However, because of the intrinsic  Lorentz symmetry  breakdown which accompanies these theories, to  physically validate them one has to demonstrate that Lorentz invariance emerges at low energies. Renormalization group methods are in general  employed  in such endeavors~\cite{Iengo}.

In this work we will study in $3+1$ dimensions  a class of models with quartic self-interactions of spinor fields. In the development of field theory this class occupies a  prominent position both in its conceptual aspects as well as in its applications. Thus, the Thirring or the Gross-Neveu model in two space-time dimensions introduced new concepts as anomalous dimension, fermion-antifermion bound states and Wilson short distance expansions. Above two dimensions these models are not renormalizable in the usual sense (in 2+1 dimensions they are renormalizable only in the context of the $1/N$ expansion). In spite of that, Nambu-Jona-Lasinio like models \cite{Nambu} have been used in four dimensions as effective theories to investigate the chiral symmetry breakdown, with cutoffs to control the ultraviolet divergences. Horava-Lifshitz like four fermion models in 2+1 dimensions were analyzed in \cite{Lima}. Some studies in four dimensions with $z=3$ have also been realized \cite{Alexander}.  Here
we extend these studies in four dimensions by considering  a   class of models
with dynamical critical exponents $z=2$ and $z=3$. Thus we will study the 
model described by the Lagrangian density
\begin{eqnarray}
 {\cal L}= \bar\psi\left[i\gamma^0 \partial_0+b_{z}(i\gamma^i\partial_i)^z-m^{z}\right]\psi+ \frac{g_\sigma}{2N}(\bar \psi \psi)^2-\frac{g_{\sigma_{5}}}{2N}(\bar \psi \gamma_{5}\psi)^{2}\label{Dirac}
\end{eqnarray}
where $\psi $ denotes a $N$-tuple of  four-component fermion fields.
As the field $\psi$ has  mass dimension $3/2$, the self-interactions above are perturbatively non-renormalizable if $z<3$. However, as we will shortly show,  with some caveats, they are renormalizable for $z=2$ in the context of the $1/N$ expansion and dimensional regularization. 
 
 For $g_{\sigma_{5}}=0$, $m=0$ and $z=1$ or $3$, Eq. (\ref{Dirac}) describes the
 usual Gross-Neveu model and one of its  higher spatial derivative extensions. They are invariant under the discrete transformation $\psi\rightarrow \gamma_{5}\psi$. This symmetry is explicitly broken by the mass term $m^{z}\bar{\psi}\psi$ and also by the kinetic  part of the Lagrangian if $z=2$.
 For $g_{\sigma}=g_{\sigma_{5}}$, $m=0$ and $z=1$ we have the chiral Gross-Neveu model which is invariant under the chiral transformation
 $\psi\rightarrow e^{i\alpha\gamma_{5}}\psi$. This symmetry is also shared by the model with $z=3$ but it is broken if $m\not= 0$ or if $z=2$. Actually, the model with $z=2$ is invariant only on the anisotropic scaling.
 
The $1/N$ expansion of these models is more easily generated by rewriting  the above Lagrangian as
\begin{equation}
 {\cal L} = \bar\psi\left[i\gamma^0 \partial_0+b_z(i\gamma^i\partial_i)^z\right]\psi-\sigma \bar \psi \psi-\frac{N}{2g_{\sigma}}\sigma^2-\sigma_{5}\bar \psi i\gamma_{5}\psi-\frac{N}{2g_{\sigma_{5}}}\sigma_{5}^2,\label{auxiliary}
\end{equation}
where henceforth we have taken $m=0$.
By the use of the equations of motion for the auxiliary scalar  fields $\sigma$ and $\sigma_{5}$,  we may reobtain Eq. (\ref{Dirac}) with $m=0$. In this new form of the Lagrangian the chiral symmetry, which holds for $g_{\sigma}=g_{\sigma_{5}}$, corresponds to the transformation  $\psi\rightarrow {\rm e}^{i\gamma_{5}\theta}\psi$ and
\begin{eqnarray}
\left(\begin{array}{c}\sigma\\
\sigma_{5}\end{array}\right)\rightarrow\left(\begin{array}{cc}\cos(2\theta) & \sin(2\theta)\\-\sin(2\theta) &\cos(2\theta)\end{array}\right)\left(\begin{array}{c}\sigma\\
\sigma_{5}\end{array}\right).
\end{eqnarray}

 Let $\sigma_{0}$ and $\sigma_{50}$ be the vacuum expectation values (v.e.v.) of $\sigma$ and $\sigma_{5}$, respectively. If any of them is nonvanishing the discrete symmetry $\psi\rightarrow\gamma_{5} \psi$ is explicitly broken.
 By shifting the  fields $\sigma $ and $\sigma_{5}$ so that $\sigma\rightarrow \sigma+ \sigma_{0}$ and $\sigma_{5}\rightarrow \sigma_{5}+ \sigma_{50}$, the Lagrangian density becomes
  \begin{eqnarray}
 {\cal L} &=& \bar\psi\left[i\gamma^0 \partial_0+b_z(i\gamma^i\partial_i)^z-\sigma_{0}-i \sigma_{50}\gamma_{5}\right]\psi-\sigma \bar \psi \psi-\frac{N}{2g_{\sigma}}(\sigma+\sigma_{0})^{2}\nonumber\\
 &&-\sigma_{5}\bar \psi i\gamma_{5}\psi-\frac{N}{2g_{\sigma_{5}}}(\sigma_{5}+\sigma_{50})^2,
 \end{eqnarray}
 furnishing the free propagator ($\not\! k\equiv k^i \gamma_i$)
 \begin{equation}
S[k]= i\frac{\gamma_{0}k_{0}-i\sigma_{50}\gamma_{5}+b_{z} (\not\!k)^z+\sigma_{0}}{k_{0}^{2}-\sigma_{50}^{2}-b_{z}^{2}k^{2z}-\sigma_{0}^{2}+i\epsilon},\label{9}
\end{equation}
for $z=1$ or $3$  and 
\begin{equation}
S[k]= i\frac{\gamma_{0}k_{0}-i\sigma_{50}\gamma_{5}+ w}{k_{0}^{2}-\sigma_{50}^{2}-w^2+i\epsilon},\label{10}
\end{equation}
with $w=b_{2}{\vec k}^2+ \sigma_{0}$, for $z=2$. 
Thus, for $z=1$ or $3$, the condition that the new auxiliary fields have vanishing v.e.v. gives the tadpole equations
\begin{eqnarray}
&-i N \,\frac{\sigma_{0}}{g_{\sigma}}&- 4 N\sigma_{0} I[\sigma_{0},\sigma_{50};z] =0,\nonumber\\
&-i N \,\frac{\sigma_{50}}{g_{\sigma 5}}&- 4 N \sigma_{50} I[\sigma_{0},\sigma_{50};z]=0 ,\label{4}
\end{eqnarray}
where
\begin{equation}
I[\sigma_{0},\sigma_{50};z] =\int\frac{dk_{0} d^{3}k}{(2\pi)^4}\frac{1}{k_{0}^{2}-\sigma_{50}^{2}-b_{z}^{2}k^{2z}-\sigma_{0}^{2}+i\epsilon}.
\end{equation}
The integral  above  can be easily calculated. We first integrate on $k_{0}$ and afterwards
promote the remaining 3-dimensional integral to $d$ dimensions.
For $z=3$ and $z=1$, respectively, employing dimensional regularization in the spatial part, we have
 \begin{eqnarray}
&I[\sigma_{0},\sigma_{50};3] &=\int\frac{dk_{0} d^{d}k}{(2\pi)^{(d+1)}}\frac{\mu^{3-d}}{k_{0}^{2}-\sigma_{50}^{2}-b_{3}^{2}k^{6}-\sigma_{0}^{2}+i\epsilon}=- \frac{i\pi}{2 b_{3}^{d/3}}\int \frac{d^d k}{(2\pi)^d} \frac{\mu^{3-d}}{\sqrt{k^6+\sigma_{0}^{2}+\sigma_{50}^{2}}}\nonumber\\
&&=\frac{i}{4 \pi^2 b_{3}}\frac{1}{d-3}+\frac{i}{24 \pi^2 b_{3}}\ln\left(\frac{\sigma_{0}^{2}+\sigma_{50}^{2}}{\mu^6}\right) 
\end{eqnarray}
and
\begin{eqnarray}
&I[\sigma_{0},\sigma_{50};1] &=\int\frac{dk_{0} d^{d}k}{(2\pi)^{(d+1)}}\frac{\mu^{3-d}}{k_{0}^{2}-\sigma_{50}^{2}-b_{1}^{2}k^{2}-\sigma_{0}^{2}+i\epsilon}=- \frac{i\pi}{2 b_{1}^{d}}\int \frac{d^d k}{(2\pi)^d} \frac{\mu^{3-d}}{\sqrt{k^2+\sigma_{0}^{2}+\sigma_{50}^{2}}}\nonumber\\
&&=\frac{-i(\sigma_{0}^{2}+\sigma_{50}^{2})}{8 \pi^2b_{1}^{3}(d-3)}-\frac{i(\sigma_{0}^{2}+\sigma_{50}^{2})}{16\pi^2 b_{1}^3}\ln\left(\frac{\sigma_{0}^{2}+\sigma_{50}^{2}}{ \mu^2}\right),
\end{eqnarray}
where, to simplify the final results, we have redefined the renormalization spot, above designated by $\mu$, to absorb some finite constants and terms that vanish when $d=3$ have been neglected.
 The divergences in the above expressions may be eliminated by conveniently defining  the  renormalized coupling constants.  For the chiral model we adopt the same counterterm so that
\begin{eqnarray}
\frac{1}{g_{\sigma R}}&=& \frac{1}{g_{\sigma}}+\frac{1}{\pi^2 b_3}\frac{1}{d-3},\nonumber\\
\frac{1}{g_{\sigma_{5 R}}}&=& \frac{1}{g_{\sigma_5}}+\frac{1}{\pi^2 b_3}\frac{1}{d-3},\label{1}
\end{eqnarray}
and therefore we can choose
\begin{equation}
\frac{1}{g}\equiv\frac{1}{g_{\sigma R}}=\frac{1}{g_{\sigma_{5 R}}}=-\frac{1}{6\pi^2 b_{3}}\ln\left(\frac{\sigma_{0}^{2}+\sigma_{50}^{2}}{\mu^6}\right). \label{5}
\end{equation}
Assuming that $b_{3}$ does not depend on $\mu$, the invariance of this result under the renormalization group,
\begin{equation}
\left(\mu \frac{\partial\phantom a}{\partial \mu}+ \beta\frac{\partial\phantom a}{\partial g}\right)\rho=0,
\end{equation}
where $\rho \equiv \sqrt{\sigma_{0}^2+\sigma_{50}^2}$, fixes $\beta=-\frac{g^2}{\pi^2 b_{3}}$, so that the model is asymptotically free which agrees with \cite{Alexander}.

One important consequence of Eqs.(\ref{4}) is the cancellation of  divergences in the two point functions of the auxiliary fields. Indeed, up to one loop,
\begin{eqnarray}
\Gamma_{\sigma \sigma} (p)&=& - \frac{iN}{g_{\sigma}}+ \int\frac{dk_{0}d^{3}k}{(2\pi)^{4}}{\rm Tr}[S(k) S(k+p)],\nonumber\\
\Gamma_{\sigma_{5}\sigma_{5}}(p)&=& - \frac{iN}{g_{\sigma_{5}}}- \int\frac{dk_{0}d^{3}k}{(2\pi)^{4}}{\rm Tr}[S(k) \gamma_{5} S(k+p)\gamma_{5}]
\end{eqnarray}
and
\begin{eqnarray}
\Gamma_{\sigma_{5} \sigma} (p)&=& i\int\frac{dk_{0}d^{3}k}{(2\pi)^{4}}{\rm Tr}[S(k) \gamma_{5} S(k+p)].
\end{eqnarray}

For $z=3$ the would be (logarithmic) divergences of these expressions are in fact absent and we obtain
\begin{eqnarray}
\Gamma_{\sigma\sigma}(0) &=& -8 N \int \frac{dk_{0}d^{3}k}{(2\pi)^{4}}\frac{\sigma_{0}^{2}}{(k_0^2 -b_{3}^{2}k^6-\sigma_{50}^{2}-\sigma_{0}^{2})^2}=-\frac{Ni}{3\pi^2 b_{3}}\frac{\sigma_{0}^{2}}{\sigma_{0}^{2}+\sigma_{50}^{2}},\nonumber\\
\Gamma_{\sigma_{5}\sigma_{5}}(0) &=& -8 N \int \frac{dk_{0}d^{3}k}{(2\pi)^{4}}\frac{\sigma_{50}^{2}}{(k_0^2 -b_{3}^{2}k^6-\sigma_{50}^{2}-\sigma_{0}^{2})^2}=-\frac{Ni}{3 \pi^2 b_{3}}\frac{\sigma_{50}^{2}}{\sigma_{0}^{2}+\sigma_{50}^{2}},\nonumber\\
\Gamma_{\sigma_{5}\sigma}(0) &=&-8 N\int \frac{dk_{0}d^{3}k}{(2\pi)^{4}}\frac{\sigma_{0}\sigma_{50}}{(k_0^2 -b_{3}^{2}k^6-\sigma_{50}^{2}-\sigma_{0}^{2})^2}=-\frac{Ni}{3 \pi^2 b_{3}}\frac{\sigma_{0}\sigma_{50}}{\sigma_{0}^{2}+\sigma_{50}^{2}}.\label{8}
\end{eqnarray}

Similarly, for $z=1$ the highest divergence (cubic) is cancelled but a logarithimic divergence still persists. To eliminate this remaining divergence,  the bare Lagrangian should contain  kinetic terms for the auxiliary fields but this can not sustain since it would turn the model indistinguishable from the Yukawa model. For this reason, in this case the model is no longer renormalizable.

The effective potential for the $z=3$ model may be obtained by integrating the one-point function for the auxiliary field $\sigma$ which gives
\begin{eqnarray}
V[\sigma,\sigma_{5}]/N&=&\frac{\sigma^{2}}{2 g_{\sigma}}- 4 i  \int (d\sigma) \sigma I[\sigma,\sigma_{5};3]+ f[\sigma_{5}]=\frac{\sigma^{2}}{2 g_{\sigma}}+\frac{1}{2 \pi^2 b_{3}}\frac{\sigma^{2}}{d-3}\nonumber\\
& +&\frac{(\sigma^2+\sigma^{2}_{5})}{12\pi^2b_{3}}\ln\frac{(\sigma^{2}+\sigma_{5}^{2})}{\mu^{6}}-\frac{\sigma^{2}}{12\pi^2 b_{3}}
+ f[\sigma_{5}],
\end{eqnarray}
where the function $f[\sigma_{5}]$  is fixed by imposing  that the derivative of $V$ with respect to $\sigma_{5}$ coincides with the one point  function of $\sigma_{5}$. Proceeding in this way we find
\begin{equation}
f[\sigma_{5}]=\frac{\sigma_{5}^{2}}{2 g_{\sigma 5}}+\frac{1}{2\pi^2b_{3}}\frac{\sigma_{5}^{2}}{d-3}-\frac{\sigma_{5}^{2}}{12\pi^2b_{3}},
\end{equation}
 resulting that
 \begin{eqnarray}
V[\sigma,\sigma_{5}]/N&=&\frac{\sigma^{2}}{2 g_{\sigma}}+\frac{\sigma^{2}_{5}}{2 g_{\sigma 5}}+\frac{1}{2 \pi^2 b_{3}}\frac{\sigma^{2}+\sigma_{5}^{2}}{d-3}\nonumber\\
& +&\frac{1}{12\pi^2b_{3}}\left[(\sigma^2+\sigma^{2}_{5})\ln\frac{(\sigma^{2}+\sigma_{5}^{2})}{\mu^{6}}-\sigma^{2}-\sigma_{5}^{2}\right] .
\end{eqnarray}
Thus, if one adopts the renormalizations (\ref{1}) and (\ref{5}) it follows that for $z=3$
\begin{equation}
V[\sigma,\sigma_{5}]/N=\frac{(\sigma^{2}+\sigma_{5}^{2})}{12 \pi^2 b_{3}}\ln\left(\frac{\sigma^{2}+\sigma_{5}^{2}}{\sigma_{0}^{2}+\sigma_{5 0}^{2}}\right)-\frac{1}{12\pi^2 b_{3}}(\sigma^{2}+\sigma_{5}^{2}).
\end{equation}
By computing the second order derivatives, we may check that the system (\ref{8})  is obtained.

 Let us now consider the case  $z=2$. Here, the term with higher spatial derivatives, $\bar{\psi}(i\gamma^{i}\partial_{i})^{2}\psi$, breaks chiral symmetry and only
the anisotropic scale invariance remains. 
This last symmetry is also broken if either
$\sigma$ or $\sigma_{5}$ or both acquires a nonvanishing v.e.v..

Proceeding as before, let again  $\sigma_{0}$ and $\sigma_{50}$ be the v.e.v. of $\sigma$ and $\sigma_{5}$,  respectively. The free fermion  propagator is given by Eq. (\ref{10}) and the analogues of Eqs. (\ref{4}) are
\begin{eqnarray}
- i\frac{\sigma_{0}}{g_{\sigma}}&+& i\int
\frac{dk_{0}d^{3}\vec k}{(2\pi)^{4}}{\rm Tr} S(k)=-i\frac{\sigma_{0}}{g_{\sigma}}-4\int
\frac{dk_{0}d^{3}\vec k}{(2\pi)^{4}} \frac{w}{k_{0}^{2}-\sigma_{05}^{2}-w^{2}+i\epsilon}=0,\\
- i\frac{\sigma_{50}}{g_{\sigma5}}&+& i\int
\frac{dk_{0}d^{3}\vec k}{(2\pi)^{4}}{\rm Tr}[ i\gamma_{5}S(k)]=-i\frac{\sigma_{50}}{g_{\sigma5}}-4 \sigma_{50}\int\frac{dk_{0}d^{3}\vec k}{(2\pi)^{4}} \frac{1}{k_{0}^{2}-\sigma_{05}^{2}-w^{2}+i\epsilon}=0.\nonumber
\end{eqnarray}

By integrating on $k_{0}$ we obtain
\begin{eqnarray}
-\frac{\sigma_{0}}{g_{\sigma}} &+&\frac{1}{4\pi^3} \int d^3k \frac{w}{(\sigma_{05}^{2}+ w^{2})^{1/2}}=0,\nonumber\\
- \frac{\sigma_{50}}{g_{\sigma5}}&+& \frac{\sigma_{50}}{4\pi^3} \int d^3k \frac{1}{(\sigma_{05}^{2}+ w^{2})^{1/2}}=0.\label{6}
\end{eqnarray}
 
We may now envisage various possibilities.
 Firstly, if $\sigma_{50}=0$ then $\sigma_{0}=0$ and reciprocally. In fact, if $\sigma_{50}=0$ the second equation in Eq. (\ref{6}) is automatically
satisfied whereas the integral  in the first equation vanishes in the
context of dimensional regularization implying that $\sigma_{0}=0$.
Reciprocally, if $\sigma_{0}=0$ the first equation dimensionally regularized also implies that $\sigma_{50}=0$. One should emphasize that these results are strictly dependent on
the absence of a term, $b_{1}\bar \psi i\gamma^{i}\partial_{i}\psi$, linear in the spatial derivatives.

For  general  nonvanishing $\sigma_{0}$ and $\sigma_{05}$,  the elliptic integrals in the system
of equations (\ref{6})  may not be expressed in terms of 
simple functions. However, by assuming that $\sigma_{50}$ is small we may go on with our analysis developing the integrals in (\ref{6})  up to second order in $\sigma_{50}$. We have
\begin{equation}
\int d^{3}k \frac{w}{(\sigma_{50}^{2}+w^{2})^{1/2}}\approx \int d^{3}k\left(1 -\frac{\sigma_{50}^{2}}{2 w^{2}}\right)= a_1 -\frac{\pi^{2}\sigma_{50}^{2}}{2\sigma_{0}^{1/2}b_{2}^{3/2}}
\end{equation}
 and
 \begin{equation}
 \int d^{3}k \frac{1}{(\sigma_{50}^{2}+w^{2})^{1/2}}\approx \int d^{3}k\left(\frac{1}{w} -\frac{\sigma_{50}^{2}}{2 w^{3}}\right)= a_2-\frac{\pi^{2}\sigma_{50}^{2}}{8\sigma_{0}^{3/2}b_{2}^{3/2}},
 \end{equation}
where $a_1$ and $a_2$ are the formally divergent integrals
 \begin{equation}
 a_1=\int d^{3}k = \frac{4}{3}\pi \Lambda^{3} 
 \end{equation}
and 
\begin{equation}
a_2= \int d^{3}k \frac{1}{w}= \frac{4 \pi \Lambda}{b_{2}}-\frac{2 \pi^{2}\sigma_{0}^{1/2}}{b_{2}^{3/2}},
\end{equation}
with $\Lambda$ being an ultraviolet cutoff.  (If, instead of the cutoff, the integrals are dimensionally regularized then the terms containing $\Lambda$ must be deleted, i.e.,
$a_1\rightarrow 0$ and $a_2\rightarrow -\frac{2 \pi^{2}\sigma_{0}^{1/2}}{b_{2}^{3/2}}$).
Using these results we may obtain the effective potential, which is given by
\begin{equation}
V_{eff}/N=\frac{\sigma^{2}}{2g_{\sigma}}-\frac{\Lambda^{3}}{3 \pi^{2}b_{2}^{3/2}}\sigma+\frac{\sigma_{5}^{2}}{2}\left(\frac{1}{g_{\sigma_{5}}}-\frac{\Lambda}{\pi^{2}b_{2}}\right)+\frac{\sigma^{1/2}\sigma _{5}^{2}}{4 \pi b_2^{3/2} }+\frac{\sigma _{5}^{4}\,\sigma^{-3/2}}{128 \pi b_2^{3/2} }\cdot 
\end{equation}

Instead of expanding in $\sigma_{50}$, another procedure consists in introducing  the term $\Delta {\cal L}= b_{1}\bar \psi i\gamma^{i}\partial_{i}\psi$ in the Lagrangian, with $b_{1}$ small  so that it could be treated perturbatively as we will discuss.  Assuming that $\sigma_{50}=0$, a great simplification is achieved. In zeroth order in $b_{1}$, the fermion propagator is 
 \begin{eqnarray}
<\psi(k)\bar{\psi}(-k)>&=&i\frac{\gamma^0k_0+\omega}{k^2_0-\omega^2+i\epsilon}\nonumber\\
&=&\frac{iP_+}{k_0-\omega+i\epsilon}-\frac{iP_-}{k_0+\omega-i\epsilon},\label{freeprop}
\end{eqnarray}
where $\omega=b_{2}\vec{k}^2+\sigma_{0}$ and $P_{\pm}=\frac{1\pm\gamma_0}{2}$ are orthogonal projectors. Because of this form of the propagator, the analytic expression for the integrand associated to a closed fermionic loop having at its vertices just matrices which commute with $\gamma_{0}$,  may be expressed as a sum of terms which have poles either in the upper or in the lower part of the complex plane of the integration variable $k_{0}$. Of course, such expression vanishes upon integration  over $k_{0}$. Thus, for example, the two point  functions of the auxiliary fields, adopting dimensional regularization on the spatial part of the loop integration variable, are
\begin{eqnarray}
&&\Gamma_{\sigma\sigma}= -i\frac{N}{g_{\sigma}}+NTr\int \frac{dk_{0} d^{d}k}{(2\pi)^{d+1}} [S(k)S(k-p)]=-i\frac{N}{g_\sigma},\label{3}\nonumber
\\
&&\Gamma_{\sigma_{5}\sigma_{5}}=-i\frac{N}{g_{\sigma5}}- NTr\int \frac{dk_{0} d^{d}k}{(2\pi)^{d+1}} [S(k)\gamma_{5}S(k-p)\gamma_{5}]=-i\frac{N}{g_{\sigma5}}+ N I[p_0,p]
\end{eqnarray}
where
\begin{eqnarray}
I[p_0,p] &=&-2\int \frac{dk_{0} d^{d}k}{(2\pi)^{d+1}}\left[\frac{1}{(k_{0}-w(k)+i\epsilon)(k_{0}-p_{0}+w(k-p)-i\epsilon)}\right.\nonumber\\&&+\left. \frac{1}{(k_{0}+w(k)-i\epsilon)(k_{0}-p_{0}-w(k-p)+i\epsilon)}\right]\nonumber\\
&=& 2 i\int \frac{d^dk}{(2\pi)^{d}}\left[ \frac{1}{p_0+w(k)+w(k-p)}+( p_{0}\leftrightarrow -p_{0})\right].
\end{eqnarray}
By performing this straightforward integration, we get
\begin{equation}
I[p_0,p] = \frac{i}{b_{2}}\frac{\Gamma[1-d/2]}{(4\pi)^{d/2}}\left [(\frac{p^2}{4}+ \frac{p_0}{2 b_2}+\sigma_{0})^{d/2-1}+ (\frac{p^2}{4}- \frac{p_0}{2 b_2}+ \sigma_{0})^{d/2-1}\right ],
\end{equation}
which for $d=3$ gives
\begin{equation}
I[p_0,p]= -\frac{i}{4\pi b_2}\left [\sqrt{\frac{p^2}{4}+ \frac{p_0}{2 b_2}+\sigma_{0}}+ \sqrt{\frac{p^2}{4}- \frac{p_0}{2 b_2}+\sigma_{0}}\right].
 \end{equation}
Thus the corresponding propagator for large momentum decreases as
 $p^{-1}$ or $1/\sqrt{p_0}$. Taking into account this behavior, the degree of superficial divergence  for a generic
  graph $\gamma$ in the models  with nontrivial one-loop two point functions  of the auxiliary field can be calculated as   follows
 \begin{equation}
 \delta(\gamma)= 5L-2 n_F -n_{A},
 \end{equation}
 where $L$ is the number of loops and $n_F$ and $n_{A}$ are the numbers of internal fermion and auxiliary lines.
 Using now the topological identities
 \begin{eqnarray}
 L=n_F+n_A-V +1, \qquad 2n_F + N_{F}= 2 V, \qquad  2 n_{A}+ N_{A}= V,
 \end{eqnarray}
  where $N_{F}$  and $N_{A}$ are the number of  external lines associated to the fermion and auxiliary fields and $V$ is the number of vertices in $\gamma$, we obtain
\begin{equation} 
\delta(\gamma)= 5-\frac32 N_F-2 N_A,
\end{equation}
 Thus, at one loop there will be divergences that may be eliminated by a
wave function renormalization of the $\psi$ field and  renormalizations of the parameters
$\sigma_{0}$ and $b_{2}$. (Similarly, for generic $z$, using (\ref{9}), we found
$\delta(\gamma)= 3+z-\frac32 N_F-z N_A$).

\begin{figure}[ht]  
  \begin{center}
    \vbox{
      \includegraphics[width=0.18\columnwidth]{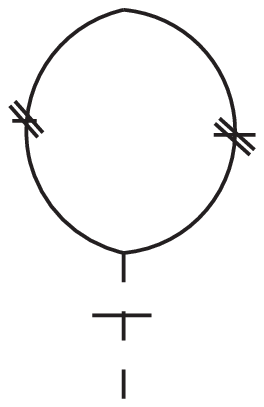}}
      \caption{Tadpole graph with two insertions of the vertex $b_{1}\bar{\psi}i\gamma^{i}\partial_{i}\psi$. The continuous and dashed lines stand for the propagators for the fermion and sigma fields.\label{Figura1}} 
      \end{center}
      \end{figure}

We still have to discuss the nonvanishing of the $\sigma$ field tadpole due to the introduction of the term linear in the spatial derivatives.
 In this new condition, up to second order in $b_{1}$, the tadpole equation becomes  (see Fig. \ref{Figura1})
\begin{eqnarray}
-\frac{i\sigma_{0}}{g_{\sigma}}&+& 2 b_{1}^{2}\int \frac{d k_{0}d^{3}\vec{k}}{(2\pi)^{4}}
\vec{k}^2 \left[ \frac{1}{(k_{0}-w+i \epsilon)^{2}(k_{0}+w-i\epsilon)}\right.\nonumber\\
&&-\left.\frac{1}{(k_{0}+w-i \epsilon)^{2}(k_{0}-w+i\epsilon)}\right]=0
\end{eqnarray}
so that
\begin{equation}
-\frac{i\sigma_{0}}{g_{\sigma}}+\frac{ib_{1}^{2}}{ b_{2}^{2}}\int \frac{d^{3}k}{(2\pi)^3}\frac{k^{2}}{(k^{2}+ \frac{\sigma_{0}}{b_{2}})^{2}}=0.\label{2}
\end{equation}

\begin{figure}[ht]  
  \begin{center}
    \vbox{
      \includegraphics[width=0.5\columnwidth]{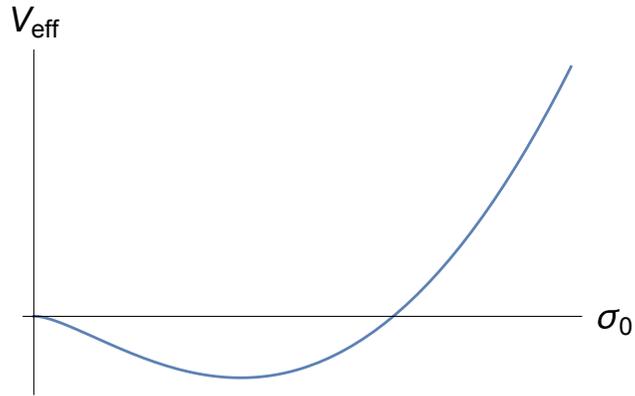}}
      \caption{General aspect of the effective potential for the one-loop effective potential in Gross-Neveu model with $z=2$.\label{Figura2}} 
      \end{center}
      \end{figure}

By using the dimensional regularization, we obtain the finite result
\begin{equation}
0=
\frac{\sigma_{0}}{g_{R}}-\frac{3 b_{1}^{2}}{8\pi b_{2}^{2}} \sqrt{\frac{\sigma_{0}}{b_{2}}}.
\end{equation}
This equation allows for $\sigma_{0}\not = 0$.
We may also calculate the effective potential by integrating the last result. However, here we.obtain it from  the effective action given by
\begin{eqnarray}
\label{efpotsigma}
\Gamma_{eff}&=&-\int d^{4}x\frac{N\sigma^{2}}{2 g_{\sigma}}- \frac{N b_{1}^{2}}{2}\int d^{4}x \int\frac{d^{4}k}{(2\pi)^{4}}{\rm Tr} [S(k) \not \! k S(k)\not \! k]=\nonumber\\
&=&-\int d^{4}x\frac{N\sigma^{2}}{2 g_{\sigma}}-2N \frac{b_{1}^{2}}{b^2_2}\int d^{4}x\int\frac{dk_0d^3k}{(2\pi)^{4}}
\frac{\vec{k}^2}{k^2_0+(\vec{k}^2+\sigma)^2},
\end{eqnarray}
where $S[k]$ denotes the fermion propagator given in Eq. (\ref{freeprop}). Assuming $\sigma$ to be constant, disregarding a factor of volume and changing the overall sign, we obtain
\begin{eqnarray}
\label{veff}
V_{eff}&=&N\frac{\sigma^2}{2 g_{\sigma}}- N\frac{b_{1}^{2}\sigma^{3/2}}{4\pi b_{2}^{5/2}}.\label{ren1}
\end{eqnarray}
One may then verify that its minimum is reached at  $\sigma=\sigma_{0}$ satisfying Eq. (\ref{2}).
For positive $g_{\sigma}$, the general behavior of this effective potential is shown in Fig. \ref{Figura2}.

 One can generalize the calculation of this effective potential for  finite temperature. To proceed in this case, we follow the Matsubara methodology, that is, we require, in the Eq. (\ref{efpotsigma}), that the zeroth component of the momentum be discrete, $k_0\to 2\pi T(n+\frac{1}{2})$, with $n$ integer, and $T$ is the temperature, and the integral over $k_0$ been replaced by the sum over $n$. After calculating the trace, one arrives at 
\begin{equation}
\label{efpotsigmaT}
V_{eff}[T]=\frac{N\sigma^{2}}{2 g_{\sigma}}+ 2N \frac{b_{1}^{2}}{b^2_2}T \sum\limits_{n=-\infty}^{\infty}\int\frac{d^3\vec{k}}{(2\pi)^3}\frac{\vec{k}^2}{4\pi^2T^2(n+\frac{1}{2})^2+(\vec{k}^2+\sigma)^2}\cdot
\end{equation}
To develop this expression, it is convenient to do first the sum and afterwards the integral, as it has been done in \cite{ourT}. We use the formula
\begin{equation}
\sum\limits_{n=-\infty}^{\infty}\frac{1}{a^2+(n+\frac{1}{2})^2}=\pi\frac{\tanh\pi a}{a},
\end{equation}
with $a^2=\frac{(\vec{k}^2+\sigma)^2}{4\pi^2T^2}$, a dimensionless parameter. 
Therefore we have
\begin{eqnarray}
\label{int}
V_{eff}[T]&=&\frac{N\sigma^{2}}{2 g_{\sigma}}+2 N \frac{b_{1}^{2}}{2b^2_2}\int
\frac{d^3\vec{k}}{(2\pi)^3}\frac{\vec{k}^2}{\vec{k}^2+\sigma}
\tanh\frac{\vec{k}^2+\sigma}{2T}.
\end{eqnarray}
We note that in the limit of zero temperature this result reproduces (\ref{efpotsigma}) after integration over $k_0$.  Here we assume that the dimensional regularization is used.

It remains now to calculate the above integral. To do it, one can introduce dimensionless variables: first of all, we replace the module of momentum $k$ as $k=\sqrt{T}t$, with $t$ being a dimensionless integration variable (remind that the mass dimension of the temperature is 2), then, introduce the dimensionless parameter $\alpha^2=\frac{\sigma}{T}$. We have
\begin{eqnarray}
V_{eff}[T]&=&V_{eff}(0)+\frac{T^{3/2}}{2\pi^2}\frac{Nb^2_1}{b^2_2}\int_0^{\infty} dt \frac{t^4}{t^2+\alpha^2}\left(
\tanh\frac{t^2+\alpha^2}{2}-1
\right),
\end{eqnarray}
 where $V_{eff}(0)$ is the effective potential at zero temperature given by  (\ref{veff}). However, this integral can be calculated only numerically.

Let us discuss our results. We have formulated a set of four-fermion Lifshitz-like models and showed, with use of the dimensional regularization, that they are power-counting renormalizable within $\frac{1}{N}$ expansion and dimensional regularization.  For $z=2$, we obtained explicitly the two-point functions of the auxiliary fields, and for $z=3$, we verified that the renormalization of the coupling constant removes also the divergence appearing in the two-point function of the auxiliary field. We discussed two possible ways to avoid vanishing of the tadpole for the model which occurs for $z=2$: firstly, we make a series expansion in the v.e.v. of the pseudoscalar auxiliary field, and secondly, we introduced a term linear in the derivatives of the fermion field but with the v.e.v. of the pseudoscalar field equal to zero  ($\sigma_{50}=0$). In all these cases we calculated the effective potential, and for $z=2$ we included the finite temperature counterpart. It is natural to expect that these results can be generalized for other values of the critical exponent, and that for all even $z$ the situations will be rather similar. Also, we note that the results for other spinor-scalar couplings, or, as is the same, for other four-fermion interactions, do not essentially differ.

\section*{Acknowledgments}
 The work by A. Yu. P. has been partially supported by the CNPq project No. 303783/2015-0, and the work by A. J. S. has been partially supported by the CNPq project No. 306926/2017-2.

\vspace*{5mm}

\end{document}